\documentclass[%
 reprint,
 amsmath,amssymb,
 aps,
 twocolumn,
]{revtex4-2}

\usepackage{printlen}
\usepackage{color,soul}
 
\usepackage{xcolor}
\usepackage{graphicx}
\usepackage{dcolumn}
\usepackage{bm}
\usepackage{hyperref}
\usepackage[mathlines]{lineno}%
\usepackage{subfigure}

\newcommand*{\affaddr}[1]{#1} 
\newcommand*{\affmark}[1][*]{\textsuperscript{#1}}
\begin{document}

\preprint{APS/123-QED}

\date{\today}

\title{The Systemic Impact of Deplatforming on Social Media}
\author{Amin Mekacher\affmark[1,$\dagger$], Max Falkenberg\affmark[1,$\dagger$,*] and Andrea Baronchelli\affmark[1,2,*]
\begin{center}
\affaddr{\affmark[1]{ \textit{City University of London, Department of Mathematics, London EC1V 0HB, (UK)}}} \\
\affaddr{\affmark[2]{ \textit{The Alan Turing Institute, British Library, London NW1 2DB, (UK)}}} \\
\affaddr{\affmark[$\dagger$]{These authors contributed equally.}}\\
\affaddr{\affmark[*]{Corresponding authors: max.falkenberg@city.ac.uk, abaronchelli@turing.ac.uk}}
\end{center}
}
\begin{abstract}
\vspace{0.5cm}
\textbf{Deplatforming, or banning malicious accounts from social media, is a key tool for moderating online harms. However, the consequences of deplatforming for the wider social media ecosystem have been largely overlooked so far, due to the difficulty of tracking banned users. Here, we address this gap by studying the ban-induced platform migration from Twitter to Gettr. With a matched dataset of 15M Gettr posts and 12M  Twitter tweets, we show that users active on both platforms post similar content as users active on Gettr but banned from Twitter, but the latter have higher retention and are 5 times more active. Then, we reveal that matched users are more toxic on Twitter, where they can engage in abusive cross-ideological interactions, than Gettr. Our analysis shows that the matched cohort are ideologically aligned with the far-right, and that the ability to interact with political opponents may be part of the appeal of Twitter to these users. Finally, we identify structural changes in the Gettr network preceding the 2023 Brasília insurrections, highlighting how deplatforming from mainstream social media can fuel poorly-regulated alternatives that may pose a risk to democratic life.
} 
\end{abstract}
\maketitle

Social media has always been controversial, with constant debate around which content should be permitted, which content should be banned, and the conditions under which a user should be deplatformed for breaking the rules \cite{owono2022_ban, unesco_hate}. Particularly since the US Capitol insurrections, the deplatforming question has become a cornerstone of the polarized public discourse, with major social media companies facing increased pressure to deplatform malicious users \cite{adl2021_qanon}. 

The rationale behind deplatforming is straightforward: Removing malicious accounts from social media helps protect other users and limits the spread of content which has the potential to cause harm \cite{jhaver2021_deplatforming, johansson2022_deplatforming}. The scientific literature supports this view showing that many harmful communities are no longer active on mainstream platforms; these groups previously thrived by posting hate speech or conspiracy theories \cite{ribeiro_2021manosphere, ribeiro_2021incel, jhaver2021_deplatforming, ali2021_deplatforming, rauchfleisch2021_deplatforming, voskresenskii2023_farright, mekacher_2022voat, russo2022understanding, evkoski2022_hate, urman2020_deplatforming, bryanov2022_deplatforming,chandrasekharan2017_ban,zhang2022_ban}, with their dense interaction networks facilitating a broad reach for their content \cite{ribeiro2018_hateful} . 

However, the benefits of deplatforming for users on a mainstream platform do not account for the impact of these moderation policies on the wider social media ecosystem. Specifically, it remains unclear how banning accounts from one platform may drive migrations to weakly regulated and poorly monitored fringe alternatives where violent narratives may develop and thrive \cite{rauchfleisch2021_deplatforming, winter2020_extremism, bovet2022_telegram,tufekci2018_social, cinelli2021_hate,naffakh2022_terrorism,romano2021_gamergate}. This is in large part because data is rarely available which permits the cross-platform tracking of social media users, particularly following account suspensions. 

In this paper, we present a unique dataset which addresses this gap, focusing on a matched cohort of users who migrated from Twitter to Gettr, a Twitter-clone that has attracted many of Twitter's most high-profile suspended accounts including US congresswoman Marjorie Taylor-Greene, media executive Steve Bannon, and conspiracy theorist Alex Jones.

Our dataset presents the near-complete evolution of Gettr from its founding in July 2021 to May 2022 including 15M posts from 785,360 active users who have posted at least once. Of these users, 6,152 are verified, 1,588 of which self-declare as active on Twitter (see Methods). For these 1,588 self-declared Twitter users with a verified Gettr account, we download their Twitter timeline from July 2021 to May 2022 totalling 12M tweets and retweets. These users represent the ``\textit{matched}'' cohort, with analysis of their Gettr posts (Twitter tweets) referred to as ``\textit{matched Gettr}'' (``\textit{matched Twitter}'') below. For the remaining verified Gettr users, we use the Twitter API to identify those accounts which have been suspended from the platform, assuming accounts share the same username on both platforms, totalling 454 accounts who constitute the ``\textit{banned}'' cohort. Finally, all remaining users who are not verified on Gettr are part of the ``\textit{non-verified}'' cohort.

In the remainder of the paper, we will overview account activity and retention on Gettr, showing that the banned cohort are 5 times more active than the matched cohort. Despite this, our results will show that these two cohorts are structurally mixed on Gettr, sharing the same politically homogenous audience and posting similar content. Using matched cohort tweets, we will show that Gettr is generally representative of the US far-right, and that matched users are more toxic on Twitter than they are on Gettr. Finally, we will highlight how Gettr had a global impact, outlining the structural changes in the Portuguese-language Gettr network that emerged in the run up to the January 2023 riots in Brazil.  

\section*{Results}

\subsection*{User acquisition and activity}

\begin{figure}[h!]
    \centering
    \includegraphics{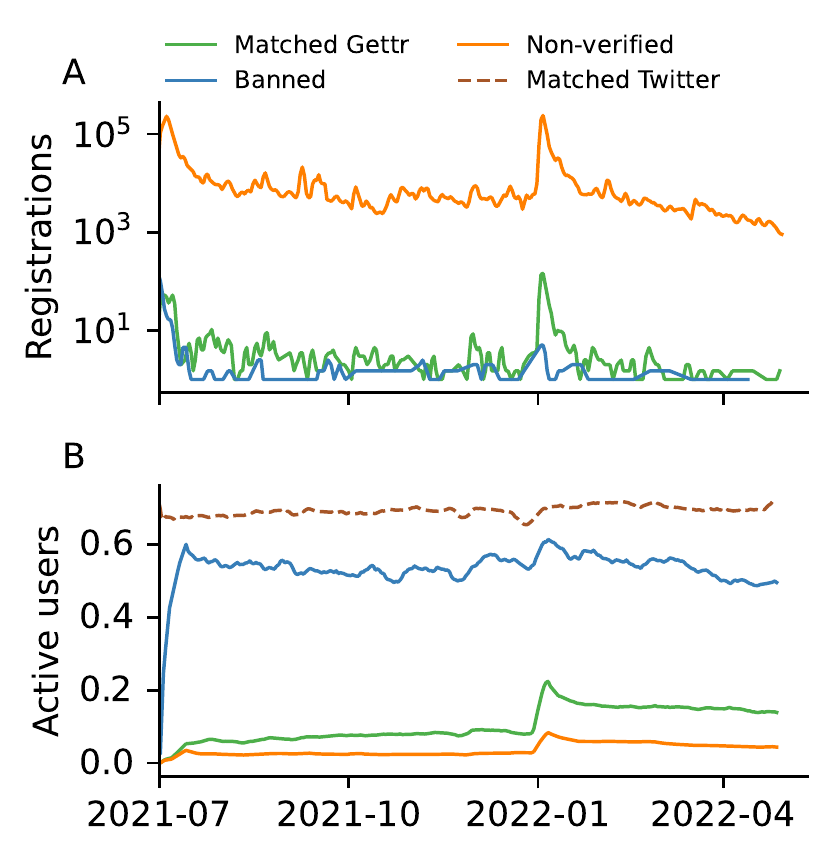}
    \caption{\textbf{User registrations and daily activity for each cohort.} (A) 3-day moving average of the daily number of users who registered on Gettr. The curve is displayed separately for the banned cohort (blue), the matched cohort (green) and other non-verified users on Gettr (orange). (B) 7-day moving average of the proportion of users from each cohort who were active on Gettr on a given day. The percentage of the matched cohort active on Twitter is also shown (dashed brown).}
    \label{fig:timeline}
\end{figure}

We start by analysing how the three cohorts of ``matched Gettr'', ``banned'' and ``non-verified'' users joined Gettr. Figure~\ref{fig:timeline}A shows that user registrations were largely steady over time with two exceptions where registrations peaked: (i) July 2021 when the platform was founded, and (ii) January 2022 following the suspension of Marjorie Taylor-Greene and Robert Malone on Twitter \cite{johnson2022_marjorie}, and the announcement by Joe Rogan that he would be opening a Gettr account \cite{mkay2022_rogan}. 

In Figure~\ref{fig:timeline}B, we show the fraction of accounts from each cohort who are active on any given day. For the matched cohort, we present their activity on both Gettr and Twitter. Focusing on the non-verified cohort, we see that a growing user base does not correlate with the growth of an engaged community, with, on average, 4\% of the non-verified cohort active on any given day. On Gettr, $10\%$ of the matched cohort are active on average, likely exceeding the value for the non-verified cohort because verified social media users are typically more active than other users \cite{deverna2022superspreader}. However, on Twitter, the matched cohort are significantly more active with $69\%$ of accounts active any given day. The activity of the matched cohort on Twitter is stable, with no evidence of a reduction in activity following the January 2022 suspensions. For the banned cohort on Gettr, activity approaches the baseline of the matched cohort on Twitter, with $53\%$ active daily, 5-times larger than for the matched cohort on Gettr, and 13-times larger than the non-verified cohort. These results are qualitatively robust if we consider exclusively English-language accounts or Portuguese-language accounts, the first and second largest Gettr demographics, respectively (see SI).

\subsection*{User retention on Gettr} 
\begin{figure*}[t]
    \centering
\includegraphics{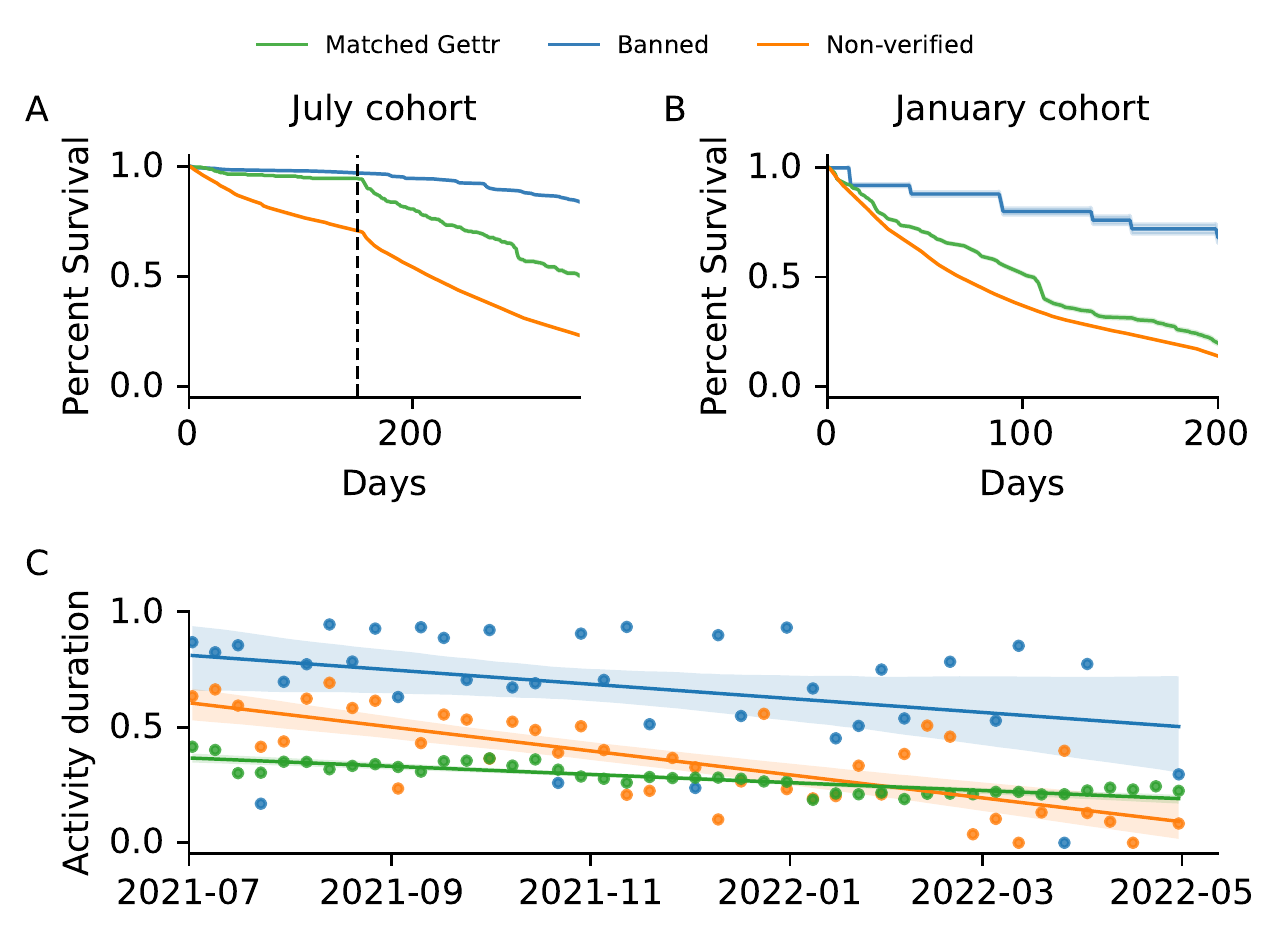}
    \caption{\textbf{User retention for key registration months and average retention by registration date over time.} (A) Kaplan-Meier survival curves for each user cohort showing the fraction of accounts who registered in July 2021 who remain active on Gettr a given number of days after registration for the banned cohort (blue), matched cohort (green) and the non-verified cohort (orange). The standard error of each curve is computed using Greenwood's formula \cite{cantor2001_greenwood} (see Methods).  The dashed line corresponds to January 1, 2022, shortly before Joe Rogan joined Gettr. 
    (B) Survival curves for January 2022. (C) Decay curves for user activity, showing the duration of their activity with respect to their registration date, normalised by the number of weeks to the end of our data collection period. Data for each cohort is fitted using linear regression ($y = ax + b$, $a = -0.007, [-0.014, 0]$, $b = 0.8, [0.65, 0.95]$ for banned users, $a = -0.011, [-0.015, -0.008]$, $b=0.6, [0.52, 0.67]$ for matched users, and $a=-0.003, [-0.004, -0.002]$, $b=0.36, [0.34, 0.37]$ for non-verified users; square
    brackets indicate 95\% confidence interval, highlighted by shaded area.)}
    \label{fig:survival}
\end{figure*}

We now focus on the retention of users on Gettr. In Fig.~\ref{fig:survival}, panels A and B show the survival curves for the proportion of users who remain active a certain number of days after registration (see Methods) for key registration months (July 2021 and January 2022 where registrations peaked, see Fig.~\ref{fig:timeline}), while panel C shows the average retention of users in each cohort over time. Survival curves for other registration months are shown in the SI and follow the same pattern with higher banned retention than matched retention. The matched cohort are consistently active on Twitter with no evidence that users stop using the platform over time: $90\%$ of the matched cohort are active in the first month covered by our dataset (07/21), and 98\% of these users remain active in our dataset's final month (04/22). This highlights that the matched cohort are established Twitter users who are committed to the platform.

Figure~\ref{fig:survival} shows that the banned cohort have the highest retention on Gettr, independent of the month in which they joined the platform, whereas the non-verified cohort and the matched cohort become inactive at a faster rate. For the highlighted registration months, we note that the January curves fall off at a sharper rate than the July curves: For the July cohort, half of the newly registered users from the non-verified cohort become idle after 216 days, compared to only 68 days for the January cohort. 

The event which clarifies these differences is the Marjorie Taylor-Greene deplatforming on Twitter. This deplatforming was denounced by Joe Rogan who opened a Gettr account on January 2, 2022, resulting in a large migration of his supporters, and supporters of Marjorie Taylor-Greene, to Gettr. However, after criticising the platform's policies \cite{friedman2022_rogan}, Rogan quit the platform on January 12. 
This ten-day period highlights how a single celebrity's endorsement resulted in a large migration to Gettr. However, the subsequent denouncement by Rogan not only resulted in many new users quitting the platform (those from the January 2022 cohort in panel B), but also resulted in many existing users quitting, see dashed line in panel A. Importantly, members of the banned cohort who registered in July 2021 did not leave Gettr at an enhanced rate after January 2022. This highlights that users who had the option to return to Twitter did so, but those who could not (due to suspension) continued to use Gettr.

Compared to previous Gettr studies which showed that users quickly become idle after registration \cite{thiel2021gettr}, possibly due to the lack of engaging content \cite{pegida2021_schwemmer}, our results reveal the discrepancy between users banned from Twitter and users who remain active on Twitter, indicating that Gettr was most successful at retaining users who had lost their Twitter audience. Our results also show that deplatforming events of exceptional prominence can induce a significant influx of accounts into a fringe platform, but not necessarily a corresponding outflux from the dominant mainstream platform. 

\subsection*{Gettr structure and content}

In order to further clarify differences between banned and matched users, we now focus on the structure and content of the Gettr social network. We start by generating a topic model using Gettr posts \cite{grootendorst2022bertopic} (see Methods). A table of topics and their description is provided in the SI. This shows that content on Gettr is dominated by issues of broad relevance to the US political right including (1) Covid-19 -- one sixth of all Gettr content, approaching one third in some months -- (2) deplatforming from Twitter and other social media platforms, (3) accusations of election fraud and the January 6th insurrection, and (4) broader issues regarding gender, abortion, gun-control, the US supreme court, and race. 

Most topics discussed on Gettr are prominent in tweets authored by the matched cohort, however, three themes are disproportionately prominent on Gettr: (i) Accusations of election fraud surrounding the 2020 US election, (ii) resistance to Covid-19 vaccine mandates, particularly in relation to the ``Freedom Convoy'' protests in Canada, and (iii) the Russian invasion of Ukraine. These are topics which are known to have been targets of the Twitter content moderation team \cite{twitter_ukraine,twitter_covid19,twitter_election}. 

We now measure whether the banned and matched cohorts are structurally segregated (or polarized) to assess whether the cohorts share the same, or different, audiences on Gettr. We measure segregation using the latent ideology, a well established method which constructs a synthetic ideological spectrum from user interactions on the platform \cite{barbera2015birds,flamino2021shifting,falkenberg2022growing} (see Methods). This measure orders the network of interactions between a set of influencer accounts (the banned and matched cohorts combined) and a set of accounts who interact with them (the non-verified cohort). By merging the banned and matched cohort into a single group, we can measure differences in how the non-verified cohort interact with banned and matched users in an unbiased manner based on purely structural factors. Note that we exclude a small number of accounts from the influencer set to avoid geographical conflation (see Methods); these are users not based in the USA. 

\begin{figure}[h!]
    \centering
    \includegraphics{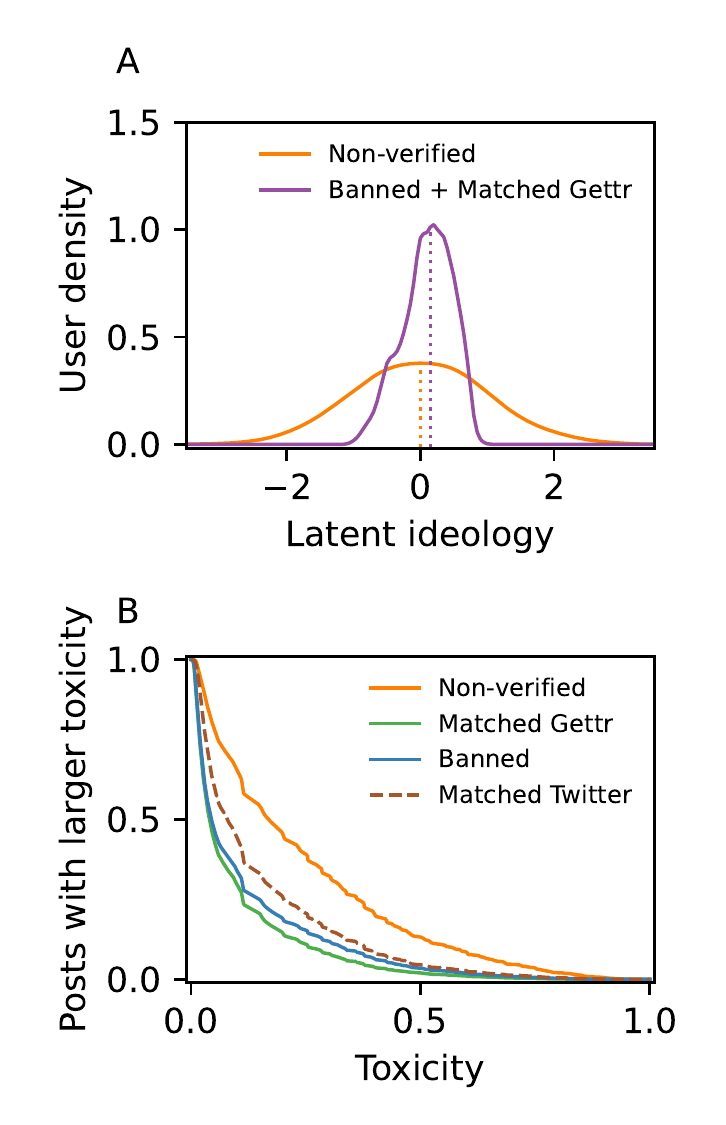}
\caption{\textbf{The latent ideology of Gettr users, and the toxicity of Gettr posts and Twitter tweets.} (A) The latent ideology is calculated using the 500 most active banned and matched users on Gettr, merged into a single influencer cohort. Unit values on the x-axis correspond to the standard deviation of the ideology distribution for all users. Both distributions are unimodal when tested using Hartigan's diptest (multimodality not statistically significant for the non-verified cohort, $p = 0.99 > 0.01$, banned and matched cohort, $p = 0.61 > 0.01$). Structural data required for calculating the latent ideology for Twitter is not available. (B) The fraction of posts (tweets) from each user cohort on Gettr (and matched Twitter) with a toxicity value larger than the value shown on the x-axis. Toxicity is calculated using the Google Perspective API \cite{perspective} (see Methods). Median toxicity [lower and upper quartile] for the non-verified cohort, $0.17 \ [0.06,0.37]$, banned cohort, $0.05 \ [0.02,0.15]$, matched cohort on Gettr, $0.04 \ [0.02,0.11]$, and matched cohort on Twitter, $0.09 \ [0.04,0.22]$.} 
    \label{fig:latent_ideology}
\end{figure}

The distribution of the latent ideology for the banned and matched cohort, and for the non-verified cohort, is shown in Fig.~\ref{fig:latent_ideology}A. Both distributions are unimodal according to Hartigan's diptest \cite{hartigan1985dip}. We observe that the banned and matched distribution falls within the bounds of the broader non-verified distribution. The banned and matched distribution is, however, significantly narrower, a feature indicative of the network centrality of these users who play a central role in the general Gettr discussion. Non-verified Gettr users are found both at the core of the Gettr discussion and at the peripheries. The central role of banned and matched users is expected since verified social media accounts typically attain higher engagement than non-verified accounts \cite{gonzalez2021bots,bovet2019_election}. 

The unimodal ideology, and the central position of the matched and banned cohorts, indicates that these users share a common audience on Gettr; segregated audiences would appear as a multi-modal ideology distribution (see examples in \cite{falkenberg2022growing,flamino2021shifting})

\subsection*{Content toxicity and twitter mentions}

We now focus on the toxicity of posts from each user cohort, shown in Fig.~\ref{fig:latent_ideology}B. Toxicity is calculated using the Google Perspective API \cite{perspective} (see Methods). The panel shows the fraction of all posts in each cohort with a toxicity score larger than the specified value. By applying a bootstrapping procedure to ensure equal sample sizes (see Methods), we find that posts authored by the non-verified cohort are more toxic than posts by the matched cohort (KS-test $D = 0.36$, $p=1.3 \times 10^{-57}$), and than posts by the banned cohort (KS-test $D = 0.32$, $p=8.0 \times 10^{-47}$). We also find that the tweets authored by the matched cohort are more toxic than Gettr posts authored by the matched cohort (KS-test $D = 0.23$, $p=8.7 \times 10^{-23}$). However, the difference between the toxicity of posts for the banned cohort and matched cohort on Gettr is not statistically significant (KS-test $D = 0.06 $, $p=0.09$).

Together, the results for the latent ideology, topic modelling, and toxicity show that, although there are significant differences in activity and retention between the banned and matched cohorts on Gettr (see Figs.~\ref{fig:timeline} and \ref{fig:survival}), there is little that distinguishes their audience and content, who merge into a single structural group. This result confirms previous research which shows that fringe platforms are politically homogeneous; platforms with this property may be referred to as ``echo-platforms'' \cite{cinelli2022conspiracy,cinelli2021echo}. In contrast, mainstream platforms are often political diverse, but with opposed political groups confined to echo-chambers \cite{cinelli2021echo,falkenberg2022growing,jamieson2008echo,barbera2020social,bovet2019_election,conover2011political,cota2019quantifying}.

Considering the toxicity of posts for each topic, we find that topics with disproportionately high toxicity are related to race (e.g., Black Lives Matter; median post toxicity [lower and upper quartile] $ = 0.40 \ [0.31, 0.52]$), focus on female US Democratic politicians ($0.38 \ [0.18,0.58]$, and discuss gender issues ($0.38 \ [0.24,0.51]$). All three topics are known to attract abusive content on social media \cite{gallagher2018divergent, mamie2021_feminist, krook2018_violence}.

We now explore possible reasons why the matched cohort are more toxic on Twitter than they are on Gettr. To do this, we analyse the Twitter accounts mentioned in tweets authored by the matched cohort. For each mentioned account, we compute the ratio between the number of users from the matched cohort who quote-tweet that account and the number of users from the matched cohort who quote-tweet or retweet that account. This ratio (referred to as the ``quote-ratio'' throughout) is instructive since there is evidence that retweets are often (but not exclusively, journalists being a known exception) used to endorse the message of the original author \cite{metaxas2015retweets,falkenberg2022growing}, whereas quote tweets allow a user to comment on a message in either a positive, negative, or neutral manner. Negative ``quoting'' behaviour is a known method of communication with ideological opponents across polarized environments \cite{guerra2017antagonism,stella2018bots}. Hence, a low quote-ratio (i.e., the account is disproportionately retweeted)  indicates general endorsement by the matched cohort of users, whereas a high quote-ratio (i.e., the account is disproportionately quote-tweeted) indicates that the matched cohort are more likely to disagree with and hold a negative view of this account. 

\begin{figure}[t]
\centering\includegraphics{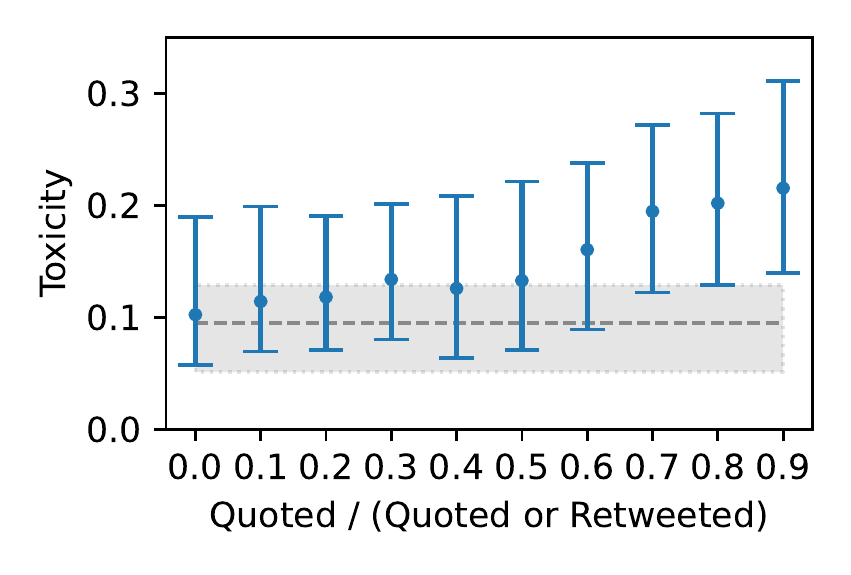}
    \caption{\textbf{Toxicity of tweets authored by the matched cohort mentioning other Twitter accounts, binned according to their quote-ratio.} The distribution of the quote-ratio is shown in Fig.~\ref{fig:qtrt_ratio}. Each point indicates the median toxicity of tweets with a quote-ratio within the binned range $[x,x+0.1)$. Error bars indicate the inter-quartile range. The dashed line indicates the median toxicity for all tweets (including those which do not mention another account) from the matched cohort, with the shaded region indicating the inter-quartile range; all data points lie above this line. }
    \label{fig:binned_toxicity}
\end{figure}

Figure~\ref{fig:binned_toxicity} shows the toxicity of tweets authored by the matched cohort, binned according to their quote-ratio. We count each mentioner-mentionee pair only once for quote-tweets and once for retweets to avoid bias from highly active accounts, and only include accounts mentioned by at least five matched users. This reveals (i) that tweets authored by the matched cohort mentioning any Twitter account are more toxic than tweets which do not mention another account, and (ii) that tweets authored by the matched cohort are more toxic if they mention an account with a high quote-ratio than if they mention accounts with lower quote-ratios. 

\begin{figure*}[t]
    \centering
    \includegraphics{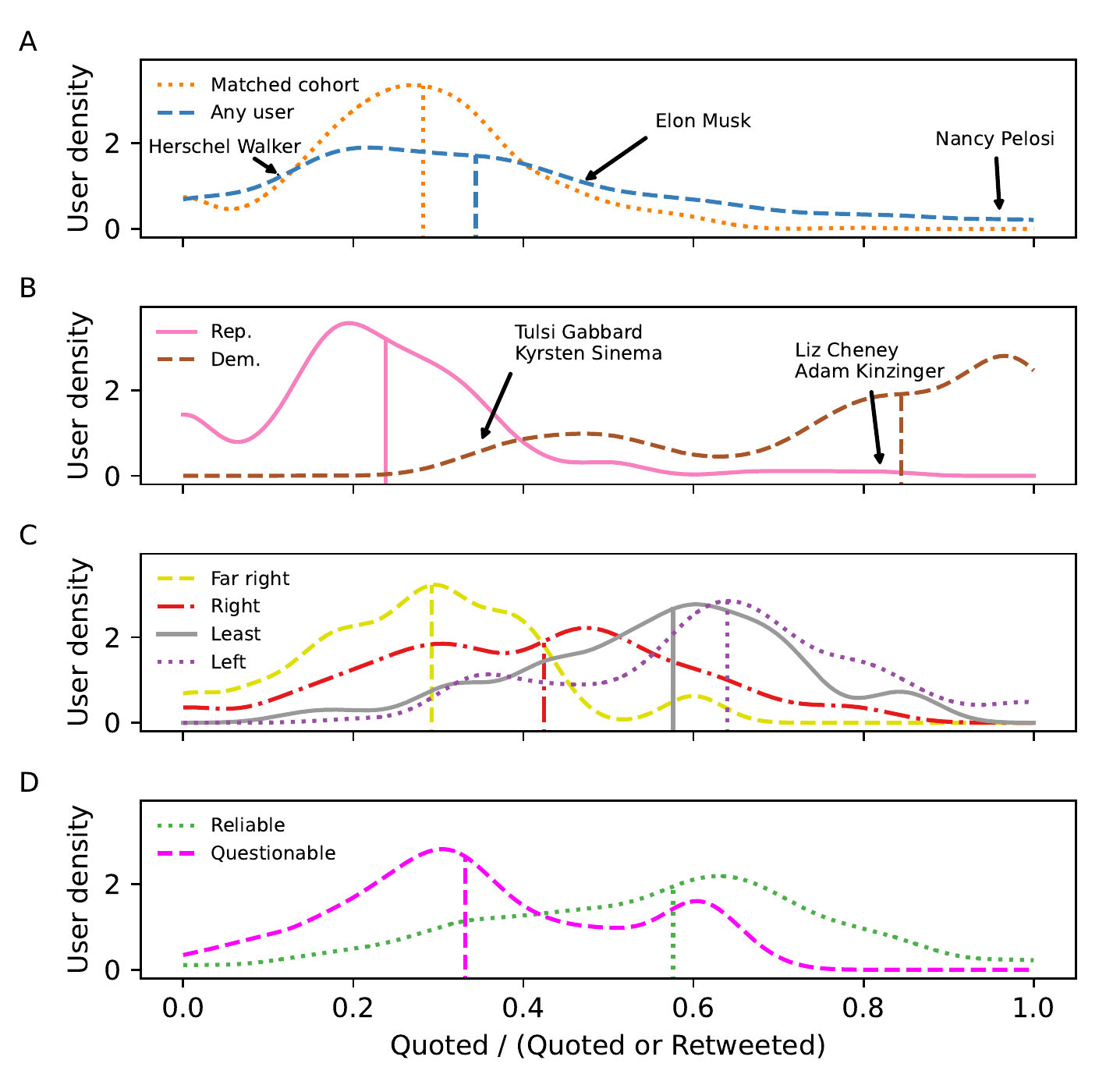}
    \caption{\textbf{The distribution of the quote-ratio of accounts mentioned on Twitter by the matched cohort.} 
    (A) The quote-ratio distribution for all mentioned accounts (blue dashed), and for mentioned accounts who are part of the matched cohort of users (i.e., a matched user mentioning another matched user; orange dotted). (B) The quote-ratio distribution for Twitter accounts belonging to known elected US Republican (pink solid) and known elected US Democrat (brown dashed) politicians. (C) The quote-ratio distribution for Twitter accounts belonging to news media organisations who have been labelled with a political leaning by MBFC. Organisations are classified as left (purple dotted), least-biased (grey solid), right (red dot-dashed), or far-right (yellow dashed). (D) The same news media organisations, but broken down according to whether they are classified as a reliable or questionable by MBFC. Vertical lines mark the median of each distribution. Annotations indicate mentioned accounts of particular interest (see text). }
    \label{fig:qtrt_ratio}
\end{figure*}

To better understand this result, we plot the distribution of the quote-ratio broken down into four groups. Figure~\ref{fig:qtrt_ratio}A shows the distribution of all users mentioned by the matched cohort, and the distribution for Twitter accounts who are also part of the matched cohort (i.e., a matched account mentioning another matched account). Three individuals are marked on the figure: (1) Republican 2022 Senate nominee Herschel Walker, the user with the lowest quote-ratio of prominent mentioned accounts ($>100$ unique mentions), (2) Democratic speaker of the house Nancy Pelosi\footnote{Speaker of the house during the timeframe of our dataset}, the user with the largest quote-ratio ($>100$ unique mentions), and (3) Elon Musk, the account with the most unique mentions. 

Figure~\ref{fig:qtrt_ratio}B shows elected US political accounts mentioned by the matched cohort, labelled using the dataset in \cite{van2020twitter}, broken down by party affiliation. This shows that Republican politicians are disproportionately retweeted (i.e., endorsed) by the matched cohort, whereas Democrats are disproportionately quote-tweeted. The individuals marked on this panel are political outliers; Liz Cheney and Adam Kinzinger are the Republican politicians with the highest quote-ratios ($>10$ unique mentions), whereas Tulsi Gabbard\footnote{Democrat during our analysis timeframe; left the Democratic party in October 2022.} and Kyrsten Sinema are the Democratic politicians with the lowest quote-ratios ($>10$ unique mentions). This shows that these politicians do not align with the dominant position of their parties. Consequently, the matched cohort are more likely to endorse the Democratic outliers, and more likely to negatively quote-tweet the Republican outliers; Liz Cheney and Adam Kinzinger have been referred to as RINOs (``Republicans in name only'') by their far-right opponents \cite{shepherd2022_cheney, fedor2021_kinzinger}.

Figure~\ref{fig:qtrt_ratio}C shows the news media organisations mentioned by the matched cohort, grouping them according to their political leaning as classified by Media Bias / Fact Check (MBFC; see Methods). Previous research confirmed that MBFC classifications are similar to classifications from other reputable media rating organisations \cite{lin2022high}.  Finally, Fig.~\ref{fig:qtrt_ratio}D repeats the analysis in panel C, but groups media outlets according to whether MBFC labels them as reliable or questionable. 

Using the distribution of all mentions (the ``any user'' curve in Fig.~\ref{fig:qtrt_ratio}A) as the baseline behaviour of the matched cohort, we find that, when tested using a two-sample Kolmogorov-Smirnov test, only the distributions of far-right media organisations in panel C (KS-test $p$-value $=0.24>0.01$; Cohen's $d=0.20$) and questionable media organisations in panel D (KS-test $p$-value $=0.29>0.01$; Cohen's $d=0.05$) are not significantly different from the baseline (see SI). The Democrat politicians distribution has the largest statistical difference to the all-mention baseline (KS-test $p$-value $=3\times10^{-16}<0.01$; Cohen's $d=2.34$). With the exception of Tulsi Gabbard, no Democratic politicians has a known Gettr account; 132 are mentioned on Twitter by the matched Gettr cohort. In contrast, 32 Republican politicians have been active on Gettr; 151 are mentioned on Twitter by the matched Gettr cohort.

Combining the evidence from the topic modelling and from the quote-ratio in Fig.~\ref{fig:qtrt_ratio} indicates that the matched cohort are aligned with the US far-right, often quote-tweeting, but not retweeting, their Democratic political opponents and moderate Republicans. In conjunction with the latent ideology in Fig.~\ref{fig:latent_ideology}, this suggests that Gettr as a whole is generally representative of the US far-right. 
These results suggest that the ability to mention one's political opponents on Twitter is part of the reason that the matched cohort are more toxic on Twitter than they are on Gettr where direct interactions with political opponents are not possible \cite{fan2021_brexit, awal2020_toxicity}. 

\subsection*{Gettr's wider impact on right-wing politics - the case of Brazil}

Journalistic reports have suggested that Gettr played a key role in facilitating the Brasília insurrection on January 8, 2023, following Jair Bolsonaro's defeat in the Brazilian Presidential elections \cite{caldeira2023_brazil, ecarma2023_brazil}. Here we investigate whether there is evidence for this role in the Gettr interaction network.

\begin{figure}[h!]
    \centering
    \includegraphics{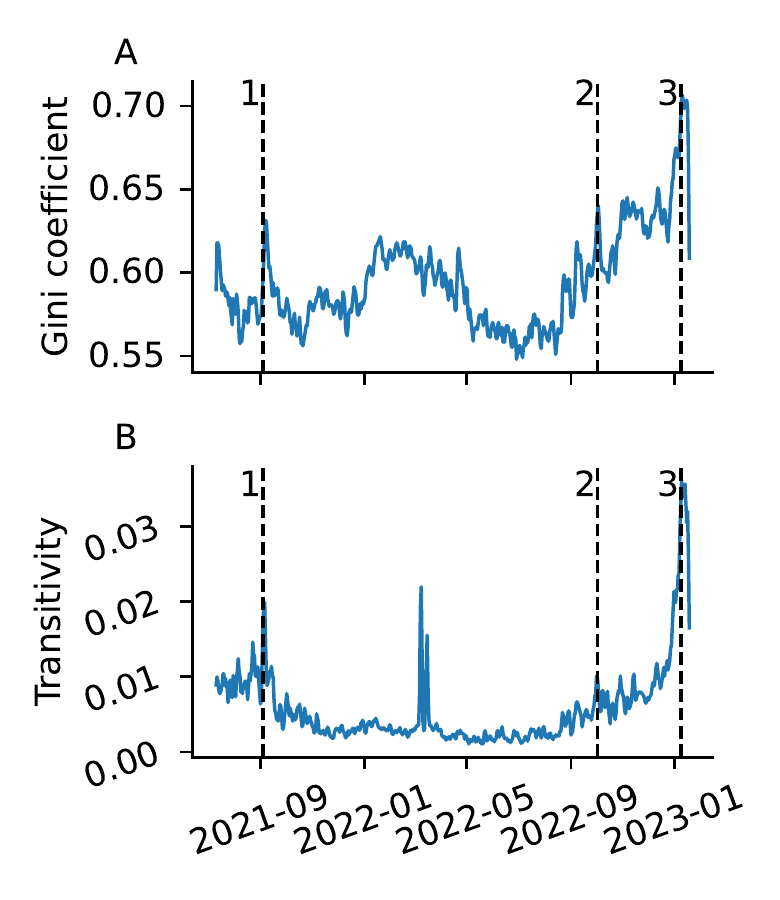}
    \caption{\textbf{Evolution of the interaction network in the Brazilian community.} Analysis of the daily interaction network, generated by considering any interaction within a 1-day window. (A) Gini-coefficient of nodes in the giant connected component. (B) Transitivity of the giant component. Dashed lines correspond to key events related to Brazilian politics and Gettr's involvement: (1) 2021 CPAC Brazil Conference, (2) the Brazilian presidential election, and (3) the Brazilian Congress attack in Brasília.}
    \label{fig:brazil}
\end{figure}

First, we study the power imbalance in the Portuguese language network by measuring the Gini coefficient of the degree distribution, shown in Fig.~\ref{fig:brazil}A. The figure shows that the Gini coefficient peaked in the run-up to the Brasília riots, which is evidence that a handful of users were responsible for shaping the collective narrative of the Portuguese language Gettr community \cite{linhong2016_inequality, guinaudeau2020_tiktok}. 
 
We now study grassroots engagement, measured by computing the transitivity of the Gettr interaction network, see Fig.~\ref{fig:brazil}B. This measure increases when a community of users densely interact with one another \cite{falkenberg2021heterogeneous,orman2013empirical}. The figure shows that network transitivity peaked during CPAC 2021, where the Bolsonaro regime and Gettr shaped their close alliance \cite{maciel2022_brazil}, and in the days leading up to the Brasília riots. Applying a Portuguese-language topic model to the network reveals that users were discussing accusations of rigged elections and claims of a corrupt media (see SI) in the lead up to the riots. 

The peak in both the Gini coefficient and the transitivity shows that leading Bolsonaro allies successfully capitalised on accusations of election fraud to generate a grassroots movement on Gettr in the wake of Bolsonaro's defeat in the Brazilian elections. These results offer new quantitative insights which build on journalistic reports of Gettr's role in the riots. Critically, our results show that even when a platform appears largely inactive, a community of idle users can be mobilised within a short time period leading to real world harms.

\section*{Discussion \& Conclusion}

In this paper, we have analysed self-declared user-level matched data to study the ban-induced platform migration from Twitter to Gettr. First, we showed that the banned cohort of users deplatformed from Twitter are more active on Gettr, and have higher platform retention, than the matched cohort who remain active on Twitter. Second, we revealed that Gettr content primarily discusses themes relevant to the US political right. Topics overepresented on Gettr are known to have resulted in account suspensions on Twitter. Third, we showed that the matched and banned cohorts share the same politically homogeneous Gettr audience. Fourth, we found that matched users are more toxic on Twitter than they are on Gettr, and that these toxic tweets often directly mention political opponents. Finally, we highlighted Gettr's broader societal impact, revealing significant structural changes in the Gettr interaction network in the run-up to the Brasília insurrections. 

The fact that the banned and matched cohorts appear similar in every regard, apart from their activity and retention on Gettr, is evidence of the systemic impact of deplatforming. 
Fringe platforms offer a safe haven where deplatformed users are free to capitalise on their supporters following suspension from Twitter. However, in this politically homogeneous environment, users are essentially confined to an ideological ``echo-platform'' \cite{cinelli2021echo,cinelli2022conspiracy} where they cannot engage and confront their political opponents. Our results hint that this ability to interact with opponents may be part of Twitter's appeal for far-right social media users, although more work is needed to fully clarify this observation. 

When users are banned from mainstream platforms, they become wholly dependent on fringe alternatives. This poses a societal risk since fringe platforms can facilitate the emergence of radical narratives and the spread of hate speech  \cite{golovchenko2022_russia, lane2021_censorship, miller2021_terrorism,rauchfleisch2021_deplatforming,bovet2022_telegram}. A lack of monitoring can, therefore, mean that signs of collective upheaval are missed. The Brasília insurrection is a clear example of this.
Critically, these considerations are not only relevant for mainstream platforms, whose policies have received the most scrutiny, but for the whole social media ecosystem. Despite branding itself as a free speech platform, Gettr has also deplatformed users for posting radical content, most notably the white supremacist Nicholas Fuentes \cite{petrizzo2021_fuentes}. 

It is important to contextualise the scope of our findings, whose limitations are avenues for future work. First, the current study only considers the migration from Twitter to Gettr, since users of other platforms do not declare their Twitter use as standard. If data becomes available, future work should extend scrutiny to other platforms.
Second, the Gettr matching feature only applies to verified users, a subset of the users who migrated from Twitter to Gettr. Analysing non-verified users migrating across platforms would clarify the differential impact of deplatforming on content creators as opposed to consumers.  
Finally, we cannot study tweets from the banned cohort since this data is not publicly available. Analysing this content would explain why some users are deplatformed, but others are not. 

Overall, our study highlights how Gettr struggles to compete with Twitter when users have free choice to use either platform. However, the decision by Twitter to deplatform a user impacts how those users view Gettr as an alternative. We anticipate that future work will build on these observations and speculate that other fringe platforms will likely show a similar dependence on their mainstream competitors. This work is urgently needed given the risks posed to democracy by poorly regulated social media \cite{tucker2017liberation,persily2020social}.

\section*{Methods}

\subsection*{Gettr data}

The data used for this study has been collected using GoGettr \cite{gogettr}, a public client developed by the Stanford Internet Observatory to give access to the Gettr API. This API allows to query user interactions, including the posts they like or share. User profiles were initially collected through a snowball sampling, by using highly popular accounts on the platform as seed users, and using the API to query their follower and following list, before repeating the same process for a random sample of the newly retrieved users. Repeating this process many times ensures that our dataset is near-complete for the studied time period.

To attract more users from Twitter, Gettr previously offered a feature that would automatically import a user's tweets upon creating an account. However, due to Twitter blocking this capability on July 10, 2021\cite{morse2021_gettr}, Gettr had to discontinue this feature. To ensure the accuracy of our results, any posts imported before July 10, 2021, and any Gettr post whose timestamp precedes the account's creation date were removed from our dataset.

To ensure our case study on the Brazilian right-wing group encompasses the Brasília insurrection, we expanded the data collection time frame for any user associated with the Brazilian community. The data collection was run in July 2022 for every user whose profile we have retrieved, and in January 2023 for the users in the Brazilian cohort.

\subsection*{Twitter data}

For each verified Gettr account where the Gettr API references their Twitter followers in the account metadata, we check that the Twitter account with the same username is active using the Twitter API (see \url{https://developer.twitter.com/en/products/twitter-api/academic-research}). Then, for each active account we download their Twitter timeline including all tweets and retweets in the period July 2021 to May 2022. This totals approximately 12 million tweets. Data was collected between September and October 2022, preceding Elon Musk's amnesty of suspended Twitter accounts.

\subsection*{User labelling}

Throughout our analysis, we label verified Gettr users as being either ``matched'' or ``banned'', depending on whether their corresponding Twitter account is active or suspended. Any verified user who decided to link their Twitter account on their Gettr page has their Twitter follower count displayed on their profile, which can also be retrieved from the Gettr API. We stress that this self-declaration permits cross-platform matching since users can ``reasonably expect'' that their Gettr accounts will be associated with their Twitter accounts.

To match accounts across platforms, we assumed that users picked the same username on both Gettr and Twitter, and we used the Twitter API to retrieve their Twitter activity. A user is classified as ``banned'' if the Twitter user endpoint returned an error indicating that the account has been suspended.

Note that for data privacy reasons, all analysis of users across platforms is aggregated at the cohort level; we do not present results for individual users. 

\subsection*{Calculating account survival}

The Kaplan-Meier estimate is a tool used to quantify the survival rate of a population (in our case, users active on a social platform) over time. For each time step $t$, we measure how many users become indefinitely inactive, and we quantify the survival rate as
\begin{equation}
    \hat{S}(t) = \prod_{t_i \leq t} \Big{(1 - }\frac{d_i}{n_i}\Big{)},
\end{equation}
where $d_i$ represents the number of users who became inactive at time $t_i$, and $n_i$ is the number of users who are still active up to time $t_i$. Greenwood's formula is used to estimate the confidence interval for the Kaplan-Meier estimate of the survivor function. For the study time $t$, the standard error is given by
\begin{equation}
    \widehat{SE}_2(t) = \sqrt{\hat{S}^2(t) \sum_{t_i \leq t} \frac{d_i}{n_i(n_i - d_i)}}.
\end{equation}

\subsection*{Topic Modelling}

Gettr and Twitter content is analysed using the BERTopic topic modelling library \cite{grootendorst2022bertopic}. This method extracts latent topics from an ensemble of documents (in our case Gettr posts and Twitter tweets). The base model uses pre-trained transformer-based language models to construct document embeddings which are then clustered. 

These methods are known to struggle with very short documents which are common on micro-blogging sites. Hence, we train our topic model using exclusively Gettr posts which are longer than 100 characters. To avoid any single user dominating a specific topic, we limit the training set to no more than 50 posts from any given user. 

\subsection*{Latent ideology}

To calculate the latent ideology on Gettr, we use the method developed in \cite{barbera2015birds,flamino2021shifting} and filtering procedures from \cite{falkenberg2022growing}. The method uses a bipartite approach where it classifies Gettr accounts as influencers or regular users. It then generates an ordering of users based on the interaction patterns of regular users with the influencer set. In the current paper, we select the matched and banned user sets as our influencers, and the remaining set of Gettr users as our cohort of regular users. 

Two factors can conflate the latent ideology: (1) account geography, and (2) a lack of user-influencer interactions. Since we are interested in the segregation of the banned and matched cohorts based on political ideology, the former is problematic because country-specific communities on social media can appear structurally segregated from a related community in other countries, even if they are politically aligned. For this reason, we remove a small number of accounts associated with the UK and China from our set of Gettr influencers. In the latter case, a lack of user-influencer interactions can be problematic since influencers with few user interactions appear as erroneous outliers when computing the latent ideology, often because they do not take active part in the conversation. Hence, we restrict our influencer set to the 500 banned and matched accounts who receive the largest number of interactions from other users on Gettr. In the current study, we consider any interaction type including comments, shares and likes. The latent ideology is robust as long as the number of influencers used is larger than 200 accounts \cite{falkenberg2022growing}. 

In order to assess the modality of the ideology distributions, we use Hartigan's diptest. This approach is used to identify polarized social media conversations and echo-chambers \cite{flamino2021shifting,falkenberg2022growing}. Hartigan's diptest compares a test distribution against a matched unimodal distribution to assess distribution modality \cite{hartigan1985dip}.

The test computes the distance, $D$, between the cumulative density of the test distribution and the cumulative density of the matched unimodal distribution. The $D$-statistic is accompanied by a $p$-value which quantifies whether the test distribution is significantly different to a matched unimodal distribution. A $p$-value of less than $0.01$ indicates a multimodal distribution.

\subsection*{Toxicity analysis}

The toxicity of Gettr and Twitter content is computed using the Google Perspective API \cite{perspective}, which has been used in several social media studies to assess platform toxicity \cite{guimaraes2020_toxicity, spika2022_toxicity, kumar2022_toxicity}. 

Given a text input, the API returns a score between 0 and 1, indicating how likely a human moderator is to flag the text as being toxic. For our analysis, we used the flagship attribute ``toxicity'', which is defined as ``[a] rude, disrespectful, or unreasonable comment that is likely to make people leave a discussion'' \cite{perspective_toxicity}.

When computing statistics for the toxicity of each user cohort, we apply a bootstrapping procedure to avoid erroneous results from variable cohort sizes. This is important  since the distribution of post toxicity is fat tailed; there are far more posts with low toxicity than high toxicity. Therefore, a smaller set of posts from a user cohort may have a lower median toxicity, purely due to sampling effects. To avoid this conflation, bootstrapping is employed where equally sized samples are drawn from each cohort (usually $100$ posts), and the median toxicity is computed for each sample. Then, the sampling procedure is repeated $100$ times to compute the median and inter-quartile range for the sampled post toxicity.

\subsection*{News media classification using Media Bias / Fact Check}

For the quote-ratio analysis in Fig.~\ref{fig:qtrt_ratio}, we identify the Twitter handles of news media outlets and classify their political leaning using Media Bias / Fact Check (MBFC; see \url{https://mediabiasfactcheck.com/}). Ratings provided by MBFC are largely similar to other reputable media rating datasets \cite{lin2022high}. 

MBFC classify news outlets under seven leaning categories: extreme left, left, center-left, least (media considered unbiased), center-right, right, extreme right. To ensure that we have enough news media outlets to enable the quote-ratio analysis, we group these classifications into four larger groups: Left -- (extreme left, left, center-left), Least -- (least), Right -- (center-right, right), Far-right -- (extreme right). Note that we have chosen to use the terminology ``far-right'' instead of ``extreme-right'' since the former is more common in the academic literature.

\subsection*{Acknowledgements}
M.F. and A.B. acknowledge the 100683EPID Project “Global Health Security Academic Research Coalition” SCH-00001-3391. M.F. thanks Alessandro Galeazzi for providing the Media Bias / Fact Check data.

\subsection*{Data availability}
Gettr and Twitter data used for this project is available in anonymised format on request at \url{https://osf.io/dx2p8/}.

\bibliography{apssamp}

\end{document}